\documentstyle[psfig,letters]{mn}

\newif\ifAMStwofonts

\newcommand{\Mo}{\mbox{$\rm M_\odot$}}
\newcommand{\Lo}{\mbox{$\rm L_\odot$}}

\newcommand{\isec}{\mbox{s$^{-1}$}}

\newcommand{\xten}[1]{\mbox{$\times 10^{#1}$}}
\newcommand{\wl}{\mbox{$\lambda$}}
\newcommand{\forb}[2]{\mbox{$[{\rm #1\, #2}]$}}
\newcommand{\ha}{\mbox{H$\alpha$}}
\newcommand{\hb}{\mbox{H$\beta$}}
\newcommand{\oiii}{\forb{O}{III}}
\newcommand{\oii}{\forb{O}{II}}

\newcommand{\nii}{\forb{N}{II}}

\ifoldfss
  \newcommand{\rmn}[1] {{\rm #1}}

  \ifCUPmtlplainloaded \else
    \NewTextAlphabet{textbfit} {cmbxti10} {}
    \NewTextAlphabet{textbfss} {cmssbx10} {}
    \NewMathAlphabet{mathbfit} {cmbxti10} {} 
    \NewMathAlphabet{mathbfss} {cmssbx10} {} 
  \fi
  \ifAMStwofonts
    \ifCUPmtlplainloaded \else
      \NewSymbolFont{upmath} {eurm10}
      \NewSymbolFont{AMSa} {msam10}
      \NewMathSymbol{\upi}     {0}{upmath}{19}
      \NewMathSymbol{\umu}     {0}{upmath}{16}
      \NewMathSymbol{\upartial}{0}{upmath}{40}
      \NewMathSymbol{\leqslant}{3}{AMSa}{36}
      \NewMathSymbol{\geqslant}{3}{AMSa}{3E}

    \fi
  \fi
\fi 

\ifnfssone
  \newmathalphabet{\mathit}
  \addtoversion{normal}{\mathit}{cmr}{m}{it}
  \addtoversion{bold}{\mathit}{cmr}{bx}{it}
  \newcommand{\rmn}[1] {\mathrm{#1}}

  \newmathalphabet{\mathbfit} 
  \addtoversion{normal}{\mathbfit}{cmr}{bx}{it}
  \addtoversion{bold}{\mathbfit}{cmr}{bx}{it}
  \newmathalphabet{\mathbfss} 
  \addtoversion{normal}{\mathbfss}{cmss}{bx}{n}
  \addtoversion{bold}{\mathbfss}{cmss}{bx}{n}
  \ifAMStwofonts
    \ifCUPmtlplainloaded \else
      \UseAMStwoboldmath
      \makeatletter
      \new@mathgroup\upmath@group
      \define@mathgroup\mv@normal\upmath@group{eur}{m}{n}
      \define@mathgroup\mv@bold\upmath@group{eur}{b}{n}
      \edef\UPM{\hexnumber\upmath@group}
      \new@mathgroup\amsa@group
      \define@mathgroup\mv@normal\amsa@group{msa}{m}{n}
      \define@mathgroup\mv@bold\amsa@group{msa}{m}{n}
      \edef\AMSa{\hexnumber\amsa@group}
      \makeatother
      \mathchardef\upi="0\UPM19
      \mathchardef\umu="0\UPM16
      \mathchardef\upartial="0\UPM40
      \mathchardef\leqslant="3\AMSa36
      \mathchardef\geqslant="3\AMSa3E
    \fi
  \fi
\fi 

\ifnfsstwo
  \newcommand{\rmn}[1] {\mathrm{#1}}

  \DeclareMathAlphabet{\mathbfit}{OT1}{cmr}{bx}{it}
  \SetMathAlphabet\mathbfit{bold}{OT1}{cmr}{bx}{it}
  \DeclareMathAlphabet{\mathbfss}{OT1}{cmss}{bx}{n}
  \SetMathAlphabet\mathbfss{bold}{OT1}{cmss}{bx}{n}
  \ifAMStwofonts
    \ifCUPmtlplainloaded \else
      \DeclareSymbolFont{UPM}{U}{eur}{m}{n}
      \SetSymbolFont{UPM}{bold}{U}{eur}{b}{n}
      \DeclareSymbolFont{AMSa}{U}{msa}{m}{n}
      \DeclareMathSymbol{\upi}{0}{UPM}{"19}
      \DeclareMathSymbol{\umu}{0}{UPM}{"16}
      \DeclareMathSymbol{\upartial}{0}{UPM}{"40}
      \DeclareMathSymbol{\leqslant}{3}{AMSa}{"36}
      \DeclareMathSymbol{\geqslant}{3}{AMSa}{"3E}
    \fi
  \fi
\fi 

\ifCUPmtlplainloaded \else
  \ifAMStwofonts \else 
    \def\upi{\pi}
    \def\umu{\mu}
    \def\upartial{\partial}
  \fi
\fi

\title[Is there really a super-massive black-hole in M87?]
{Is there really a super-massive black-hole in M87?
\thanks{Based on observations with the NASA/ESA Hubble Space Telescope,
obtained at the Space Telescope Science Institute, which is operated
by AURA, Inc., under NASA contract NAS 5-26555 and by STScI grant 
GO-3594.01-91A} }
\author[A. Marconi et al.]
       {A.~Marconi$^{1,2}$, David~J.~Axon$^{1,3,4}$, F.D.~Macchetto$^{1,3}$,
        \newauthor
        A.~Capetti$^5$, W.B.~Sparks$^1$, P.~Crane$^6$  \\
        $^1$Space Telescope Science Institute, 3700 San Martin Drive,
	    Baltimore, MD 21218, USA \\
	$^2$Dipartimento di Astronomia e Scienza dello Spazio, Universit\`a di
	    Firenze, Largo E. Fermi 5, I-50125, Italy \\
	$^3$Affiliated to the Astrophysics Division,
	    Space Science Department of ESA,
	    Space Telescope Science Institute \\
	$^4$Nuffield Radio Astronomy Laboratory, Jodrell Bank,
	    Macclesfield, Cheshire, UK \\
	$^5$Scuola Superiore di Studi Superiori Avanzati,
	    Via Beirut 2-434014 Trieste, Italy \\
	$^6$European Southern Observatory, Karl-Schwarzschild-Str. 2,
	    D-85748 Garching bei Munchen, Germany
	}

\date{Received; Accepted}

\pagerange{\pageref{firstpage}--\pageref{lastpage}}
\pubyear{1997}

\begin{document}

\maketitle

\label{firstpage}

\begin{abstract}
We present the {\it first} HST long-slit spectrum of a gaseous disk
around a candidate super-massive black-hole.
The results of this study on the kinematics of the gaseous disk in M87
are a considerable improvement in both spatial resolution and accuracy
over previous observations and requires a projected mass of 
M$_{\rmn{BH}}(\sin i)^2 = (2.0\pm 0.5)\xten{9}\Mo$
(M$_{\rmn{BH}}=3.2\xten{9}\Mo$ for a disk inclination $i=52^\circ$)
concentrated within a sphere whose radius is less than 0\farcs05 (3.5 pc)
to explain the observed rotation curve.
The kinematics of the ionized gas is well described by a thin disk in keplerian
motion.
A lower limit to the mass-to-light 
ratio of this region is M/L$_{\rmn{V}}\simeq$110, significantly strengthening
the claim that this mass is due to the presence of a central black-hole in M87.
\end{abstract}


\section{Introduction}

One of the cornerstones of the contemporary view of the physics of active
galactic nuclei is that their energy output is generated by accretion of 
material on to a massive black-hole (e.g. Blandford 1991, Antonucci 1993
and references therein).
Considerable controversy has surrounded attempts to verify the existence
of black-holes in nearby giant elliptical galaxies using ground-based 
stellar dynamical studies (Sargent et al. 1978, Young et al. 1978,
Duncan \& Wheeler 1980, Binney and Mamon 1982). To-date the
best available data remains inconclusive,
largely because of the difficulty of detecting the
high-velocity wings on the absorption lines,
which are the hallmark of a black-hole (van der Marel 1994,
Dressler \& Richstone 1990, Jarvis \& Pelletier 1991,
Kormendy \& Richstone 1995).
One of the major goals of HST has been to establish or refute the existence 
of black-holes in active galaxies by probing the dynamics of AGN at much
smaller radii than can be achieved from
the ground. Already persuasive evidence has been found that apparently quiescent
nearby galaxies contain black-holes (van der Marel 1997a,
Kormendy et al. 1996) but, locally, there are few candidate galaxies which might
harbour a super-massive black-hole large enough to have sustained Quasar
type activity earlier in their evolution (Rees 1997).
M87 is the nearest and brightest elliptical believed to harbour such
a super-massive black-hole.

M87, the dominant giant elliptical galaxy at the center of the Virgo cluster,
was the first galaxy for which tenable stellar dynamical and photometric 
evidence was advanced for the presence of a supermassive black-hole
(Sargent et al. 1978, Young et al. 1978). Subsequent stellar dynamical models
showed however that the projected stellar density and the observed rise in 
velocity dispersion did not necessarily imply the presence of a black-hole,
but could still be explained by the presence of a stellar core with an
anisotropic velocity tensor(Duncan \& Wheeler 1980, Young 1980,
Binney and Mamon 1982).

An important step in the quest to verify the existence of a massive black-hole 
in M87 has been the discovery from HST emission line (Crane et al. 1993,
Ford et al. 1994) and continuum \cite{macchetto96}
imagery of a circum-nuclear disk of ionized gas which is oriented approximately
perpendicularly to the synchrotron jet \cite{sparks}. Similar
gaseous disks have also been found in the nuclei of a number of other massive
galaxies (Jaffe et al. 1993, Ferrarese et al. 1996).
Because of surface brightness limitations on stellar dynamical studies 
at HST resolutions, the kinematics of such disks are in practice likely 
to be the only way to determine if a central black-hole
exists in all but the very nearest galaxies.

Previous HST spectroscopic observations (Harms et al. 1994,
Ford et al. 1996) of the gas disk at several discrete locations on
opposite sides of the nucleus showed a velocity difference of
$>$1000 km \isec. On the assumption that these motions arise in a thin
rotating keplerian disk, this lead to a central dark mass
of M87 in the range 1 to 3.5\xten{9}\Mo.
An important shortcoming of this estimate is the assumption about the
inclination of the disk with respect to the line of sight 
which cannot be properly determined neither by using the few velocities
derived from the FOS observations nor from the WFPC2 imaging data.
Harms et al. (1994)
assumed a disk inclination of $42^\circ\pm5^\circ$, as derived by
an ellipse fitting of the \ha+\nii\ image, and this resulted
in a misleadingly accurate mass value. Actually, such value
for the inclination was determined at distances between 0\farcs3 and 0\farcs8 
from the nucleus, because in the inner regions the disk
emission was subsumed by the bright central point-source.
Indeed Ford et al. (1996) quoted a larger uncertainty in their mass
determination but it is not clear from that short paper whether they have fully
taken into account all of the possible values of the disk inclination
which are consistent with the data.
Implicit in this measurement of the
mass of the central object is also the assumption that the gas 
motions in the innermost regions reflect keplerian rotation and 
not the effects of non-gravitational forces such as interactions with 
the jet which are known to dominate the gas motions in the inner regions of
many AGN (Whittle et al. 1988, Axon et al. 1989). 
Establishing the detailed kinematics of the disk is therefore critical.
Currently their result still provides the most convincing
observational evidence in favour of the black-hole model.

In order to determine whether the motions observed in M87 are due to keplerian
rotation, we have re-investigated the velocity field at high spatial 
resolution with HST. Our results
demonstrate that both the observed rotation curve and line profiles 
are consistent with a thin-disk in keplerian motion and allow us 
to determine an improved estimate for the mass of the black-hole.
On the assumption that the distance to M87 is 15 Mpc,
0\farcs1$\simeq$7pc.

\section{Observations and Data Reduction}

\begin{figure}
\psfig{figure=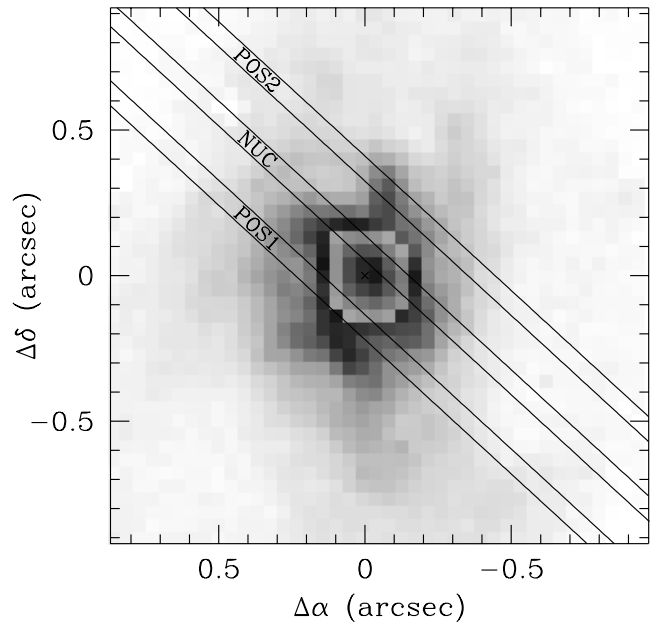,width=8.5cm,angle=-90}
\caption{\label{fig:slitpos}
The slit positions discussed in the text are marked with solid lines 
on an \ha+\nii\ image from the HST archive.
North is up and east is left.}
\end{figure}

The circumnuclear disk of M87 was observed on 1996 July 25$^{th}$,
with the COSTAR corrected Faint Object Camera f/48 long-slit spectrograph 
on board the Hubble Space Telescope.
An F305LP filter, was used to isolate the first order spectrum which 
covers the 3650-5470 \AA\ region. All the
spectra have a pixel size of 0\farcs0287 in the spatial direction and 
cover the spectral lines of \oii\wl\wl 3726,3729 \AA,
\hb\wl 4861 \AA, and \oiii\wl 4959,5007 \AA\ at a resolution of 
1.78 \AA\  per pixel and were taken in the 1024x512 non--zoomed format
of the Faint Object Camera.
An interactive acquisition image (F140W filter) was first obtained to
accurately locate the nucleus. 
The slit, with dimensions of 13\farcs5$\times$0\farcs063 at a position angle of
47$^\circ$, was then stepped across the nucleus at three positions
separated by 0\farcs2.
The integration times for the inner spectrum was 7761 second while those of the
two outer spectra where 2261 seconds.
Comparing the continuum flux observed in the spectra with the
luminosity distribution derived from a F342W FOC, f/96 archival image
we could estimate that the inner slit position was located to within
0\farcs07$\pm$0\farcs01 of the nucleus.
The slit positions used are shown in Figure \ref{fig:slitpos}
overlaid on an \ha+\nii\ WFPC2 image of the gaseous
disk of M87 obtained from the HST data archive.

The distortion  induced by the optics and the magnetic
focusing of the detector was corrected by using the 
equally spaced grid of reseau marks, etched onto the photocathode
of the detector \cite{focman}.
The distortion induced by the spectroscopic mirror and the grating
were corrected using spectra of the planetary nebula NGC 6543
and of the globular cluster 47 Tucanae for the slit
and dispersion directions respectively which are then
characterized by maximum uncertainties of 0.45 pix and 0.28 pix
(i.e. 0.8 \AA\ and 8 mas) when comparing measurements all across the detector.
After the background subtraction, the only line with enough
signal to be suitable for velocity measurements 
was \oii\wl 3726,3729 \AA, which was then fitted
(row by row) using the task LONGSLIT in the TWODSPEC FIGARO 
package \cite{figaro}. In all cases the line profile is
well represented by a gaussian function.
Full details of the observation strategy and
data reduction are given in a companion paper \cite{bigpaper}.

\section{Results}

\begin{figure}
\psfig{figure=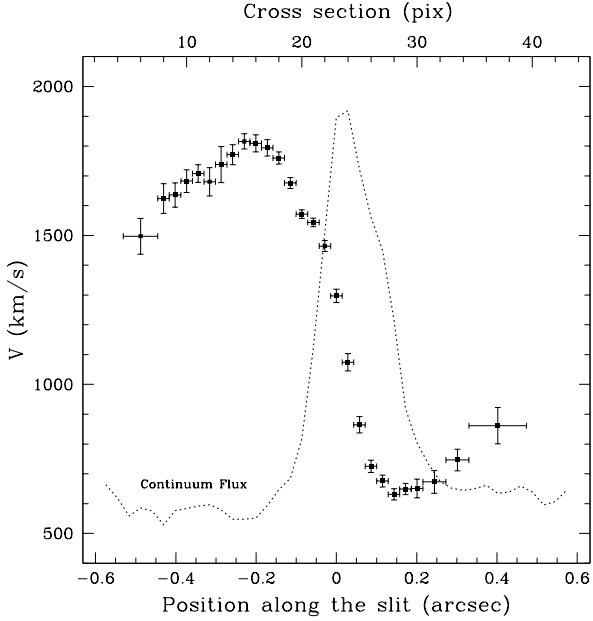,width=8.5cm,clip=yes}
\caption{\label{fig:results} 
Observed rotation curve from the \oii\wl\wl 3726,3729 \AA\ doublet
for the spectrum with the smallest impact parameter to the nucleus of 
M87 compared with the continuum flux distribution along the slit. The
uncertainties in the measured velocities and positions are 65 km \isec,
at 3727\AA, and 0.28 pixels (8 mas)
respectively. }
\end{figure}

The measured variations of radial velocity and continuum flux
for the central slit are shown in Figure \ref{fig:results}.
It is immediately apparent 
that the observed velocity field does not show the very steep rise 
in velocity one would naively expect to be the signature of a
massive black-hole.
Instead over the central $\pm$0\farcs2 the velocity
varies approximately linearly with position, with an amplitude of $\pm$600
km \isec, while at larger radii it flattens, and
eventually starts to turn over.
As shown in the following section, this apparent flattening
is a consequence of both the finite distance between the slit
and the nucleus and the smearing of the instrumental spatial
PSF.

The comparison between the new FOC data and the archival FOS observations
is shown in Fig. \ref{fig:fit} where we include only those FOS
apertures which overlapped our slit at NUC.
Within their substantially larger uncertainties, the previous
velocity measurements (Harms et al. 1994,
Ford et al. 1996) are in reasonable accord with our results.
The comparison shows clearly the increase in spatial
resolution and sampling of the FOC observations with
respect to the FOS ones. Moreover, our data provide considerably more
reliable relative spatial positions, since all the data points were obtained
simultaneously on the detector.
On the contrary, the limitations of the FOS target acquisition procedure
can produce errors as large as 0\farcs1 \cite{vandermar97s}
and such uncertainty in the position is crucial in the inner 0\farcs2
when comparing observations taken at different epochs.

\section{Discussion}

To understand the implications of our results for the mass distribution in 
the center of M87 we built models of the gas kinematics both under the 
assumption of the existence of a central black-hole,
and an extended mass distribution following a Plummer Potential.

In the hypothesis of a thin disk rotating in circular orbits, 
neglecting the slit width, instrumental PSF and luminosity
distribution of the line, the observed velocity is given by:
\begin{equation}
\label{eq:disk}
V=V_{sys} - \Phi(R)^{0.5}\frac{X}{R}\sin i
\end{equation}
where
\[
X=-b\sin\theta+s\cos\theta \mskip 20mu
Y=b\cos\theta+s\sin\theta 
\] and
\[
R=\left(X^2+Y^2\right)^{0.5}
\]
$b$ is the distance of the slit from the nucleus, $s$ is
the coordinate along the slit and $R$ represents the ``true'' distance
of each point on the slit from the nucleus ($s=0$ when $R=b$).
$V_{sys}$ is the systemic velocity, $i$ is the inclination of the
disk with respect to the line of sight ($i=90^\circ$ for the
edge-on case), $\theta$ is the angle between the slit and the line 
of nodes. $\Phi(R)$ is the gravitational potential and results
\[
\Phi(R) = \frac{GM_{BH}}{R} \mskip 20mu  \rmn{Keplerian}
\]
\[
\Phi(R) = \frac{GM}{\left(R^2+R_C^2\right)^{0.5}} \mskip 20mu \rmn{Plummer}
\]
where $M_{BH}$ and $R_C$ are the black-hole mass and core radius,
respectively.

The effects of the f/48 spatial PSF and the finite slit size 
are taken into account by averaging the velocity field using the
luminosity distribution and PSF as weights.
The model rotation curve is thus given by:
\begin{equation}
V_{ps}(S) =
\frac{\int_{S-\Delta S}^{S+\Delta S} ds \int_{B-h}^{B+h} db
\int\int_{-\infty}^{+\infty} db^\prime ds^\prime
V(s^\prime,b^\prime) I(s^\prime,b^\prime) P}
{ \int_{S-\Delta S}^{S+\Delta S} ds \int_{B-h}^{B+h} db
\int\int_{-\infty}^{+\infty} db^\prime ds^\prime
I(s^\prime,b^\prime) P }
\end{equation}
where $V(s^\prime,b^\prime)$ is the keplerian velocity derived in eq.
\ref{eq:disk},
$I(s^\prime,b^\prime)$ is the intrinsic luminosity distribution of the
line, $P=P(s^\prime-s, b^\prime-b)$ is the spatial PSF of the f/48
relay along the slit direction.
$B$ is the impact parameter (measured at the center of the slit)
and $2h$ is the slit size, S is the position along the slit at which
the velocity is computed and $2\Delta S$ is the pixel
size of the f/48 relay.
For the PSF we have assumed a gaussian with 0\farcs08 FWHM i.e.
\begin{equation}
P = \frac{1}{\sqrt{2\pi\sigma^2}} \exp \left(
-\frac{1}{2}\frac{(s^\prime-s)^2}{\sigma^2}
-\frac{1}{2}\frac{(b^\prime-b)^2}{\sigma^2} \right)
\end{equation}

\begin{figure}
\psfig{figure=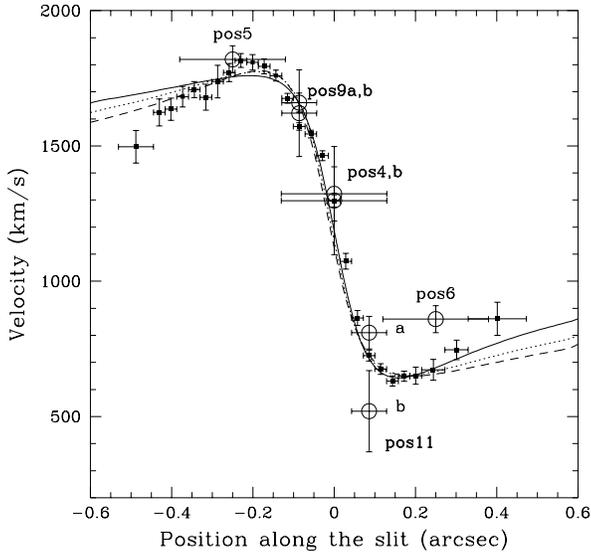,width=8.5cm}
\caption{\label{fig:fit} 
Comparison between the FOC rotation curve and the archival
FOS observations. The labels ``pos\#'' refer to the notation
used by Ford et al. (1996). ``a'' and ``b'' indicate positions
which were nominally coincident but resulted slightly shifted 
due to acquisition uncertainties. The error-bars in position
represent the aperture diameters, those in velocity represent
the dispersion of the measurements on different lines.
The solid, dotted and dashed lines are model fits which are
discussed in the text.}
\end{figure}

Because the intrinsic surface brightness
distribution of the emission line disk interior to the
HST PSF is not known, we modeled this with two extreme parameterized forms,
either as a power law or as an exponential.
We then fitted the model rotation curves using a $\chi^2$ minimization.
The free parameters of the fit, in the case of the
black-hole model, were the inclination of the disk with respect
to the line of sight, the angle between the disk and the line of nodes 
(positive angles indicate that the line of nodes has a larger position angle 
than the slit), the position along the slit of the point
closest to the nucleus, the impact parameter of the slit,
the systemic velocity and the mass of the black-hole.
The uncertainties are dominated by the poorly determined inclination 
of the gas disk within the HST PSF. 
Taking into account the possible ranges of variation of the model free
parameters, which were derived using a large Montecarlo generated 
grid of solutions, we find a best estimate of the projected mass of 
the black-hole of M$_{\rmn{BH}}(\sin i)^2 = (2.0\pm 0.5)\xten{9}\Mo$.
Three representative model fits are shown in Fig. \ref{fig:fit}.
The solid, dashed and dotted lines 
represent three black-hole model fits which are compatible with 
the observations and differ in the value of the angle between the slit 
and the line of nodes of the gaseous disk ($\theta\simeq-9^\circ$,
$7^\circ$ and $1^\circ$, respectively).
The best fit model, represented by the solid line in the upper panel
has a disk inclination,
$i=51^\circ (39^\circ\mbox{; }65^\circ)$,
angle of line of nodes, $\theta=-9^\circ (-15^\circ\mbox{; }13^\circ)$,
impact parameter, $b = 0\farcs08 (0\farcs06\mbox{; }0\farcs08)$ and a systemic
velocity, $V_{sys}=1290$ km \isec\ (1080; 1355),
leading to a mass estimate for the black-hole of 
M$_{\rmn{BH}} = (3.2\pm 0.9)\xten{9}\Mo$.
The values in parentheses represent the allowed errors on the values of 
the parameters.
As shown by this analysis, the main source of uncertainty on
the {\it deprojected} mass is the value of the disk inclination 
which more than doubles the total error.

In the case of an extended mass (Plummer)
distribution the observations are reproduced
only if at least 60\% of the total mass is confined within a sphere of 
0\farcs05 ($\simeq$3.5pc).
Since the maximum spatial extent is determined by the 
PSF FWHM (0\farcs08) this possible mass distribution is therefore 
indistinguishable from a point mass.

We have demonstrated that the data are fully compatible
with a thin disk in keplerian rotation around a central mass
condensation, at least within the inner 0\farcs2-0\farcs3.
At larger distances, the signal-to-noise in our data decreases and,
consequently, the error bars increase.
The few points that appear to deviate from the keplerian rotation
curve, do not provide enough information or weight to
warrant the fit of more complex models, such as thick disks
or sub-keplerian disks with outflows.

Can we rule out the alternative explanation that the observed rotation
is due to an extended mass?
A strong argument against this interpretation is that a mass of
$(3.2\pm 0.9)\xten{9}\Mo$ distributed within 3.5pc yields a mean density 
of $\approx2\xten{7}\Mo$ pc$^{-3}$,
higher than the highest density encountered in the collapsed cores
of galactic globular clusters.
Moreover, the V band flux in a region 16$\times$16 pc$^2$
of the nucleus yields a mass-to-light ratio
M/L$_V$$\simeq$110 \Mo/L$_{V\odot}$ (L$_{V\odot}$=0.113 \Lo) which is far
higher than those predicted for stellar clusters by evolution synthesis
\cite{bruzual}.
The mass condensation in the nucleus of M87 can therefore not be a cluster 
of normally evolved stars. Even more exotic possibilities
(e.g. massive clusters of neutron stars, other dark objects etc.) 
have been discussed by van der Marel et al. (1997a) in the case
of M32 and found to be equally improbable. Therefore the most natural
explanation for the central mass condensation of M87 is that it is due 
to the presence of a black-hole.

The data and the analysis presented by Harms et al. (1994) and
Ford et al. (1996) indicated the presence of a velocity shear which 
could be consistent which the presence of a black-hole. 
Here we have demonstrated the existence of a keplerian rotation curve
in the disk and 
we have also excluded the possibility that it might be due to a mass
concentration more extended than our spatial resolution.

\section{Conclusions}

We have presented the {\it first} HST long-slit spectrum of a
gaseous disk around a candidate supermassive black-hole.
We have obtained a rotation curve in the \oii\wl 3736,3729
which extends up to $\simeq$1\arcsec\ from the nucleus.
We have modeled the rotation curve in the case of a thin
disk in circular orbits around a supermassive black-hole
and we have derived a projected mass of
M$_{\rmn{BH}}(\sin i)^2 = (2.0\pm 0.5)\xten{9}\Mo$
(M$_{\rmn{BH}}=3.2\xten{9}\Mo$ for a disk inclination, $i=52^\circ$)
concentrated within a sphere whose radius is less than 0\farcs05 (3.5 pc).
We have shown that the observed characteristic of the data
are well explained under the working hypothesis of circular
keplerian orbits and that there is no
substantial contribution from a mass distribution
more extended than our spatial resolution.
Given the inferred mass-to-light ratio M/L$_V$$\simeq$110 \Mo/L$_{V\odot}$
the most natural and likely explanation is that of a 
supermassive black-hole.
In conclusion, with respect to previous determinations, we have 
improved the accuracy of the mass estimate, demonstrated its
reliability by verifying the assumption of the thin
disk in circular, keplerian rotation and excluded the possibility
of a mass distribution more extended than the spatial PSF.

To make further progress there are a number of possibilities the easiest 
of which is to make a more comprehensive and higher signal-to-noise 2D 
velocity map of the disk to better constrain
its parameters. The biggest limitation of the present data is that,
even by observing with HST at close to its optimal resolution at 
visible wavelengths, some of the important features of the disk
kinematics are subsumed by the central PSF.
Until a larger space based telescope becomes
available, the best we can do is to study the gas disk in Ly$\alpha$
and gain the Rayleigh advantage in resolution by moving to the UV.
This may be the only way to proceed because of the difficulty
of detecting the high velocity wings which characterize the stellar 
absorption lines in the presence of a supermassive black-hole.

\section*{Acknowledgments}

We thank M. Livio, S. Casertano and M. Stiavelli for useful discussion
through the course of this work.
We thank the referee, Dr. S. Sigurdsson, for his helpful comments on the paper.
The results are based on 
observations with the NASA/ESA Hubble Space Telescope, obtained at the 
Space Telescope Science Institute. STScI provided partial support through 
a GO grant (A.M.). STScI is operated by the Association of Universities 
for Research in Astronomy, Inc., under contract to NASA.

\vfill
\bsp
\label{lastpage}

\begin{thebibliography}{99}
%
\bibitem[Antonucci 1993]{antonucci}
Antonucci R., 1993, ARA\&A, 31, 473
%
\bibitem[Axon et al. 1989]{axon}
Axon D. J., Unger S. W., Pedlar A., Meurs E. J. A, Whittle D. M. 
and  Ward M. J., 1989, Nature, 341, 631
%
\bibitem[Binney \& Mamon 1982]{Binney}
Binney J. and Mamon S., 1982, MNRAS, 200, 361
%
\bibitem[Blandford 1991]{blandford}
Blandford R.D., 1991, ``Physics of AGN'', Proceedings of Heidelberg Conference,
Springer-Verlag, eds. W.J. Duschl and S.J. Wagner, p. 3
%
\bibitem[Bruzual 1995]{bruzual}
Bruzual G.A., 1995,  ``From Stars to Galaxies: the Impact of Stellar Physics on 
Galaxy Evolution'', Proceedings of
Crete Conference, eds. C. Leitherer, U. Fritze-von Alvensleben and J. Huchra,
ASP Conf. Series, vol. 98, p. 14
%
\bibitem[Capetti et al. 1996]{capetti}
Capetti A., Axon D.J., Macchetto F., Sparks W. B. 
and Boksenberg A., 1996, ApJ, 469, 554
%
\bibitem[Crane et al. 1993]{crane}
Crane P., Stiavelli M., King I.R., Deharveng J.M., Albrecht R., Barbieri C.,
Blades J.C., Boksemberg A., Disney M.J., Jakobsen P., 1993, AJ, 106, 1371
%
\bibitem[Dressler \& Richstone 1990]{dressler}
Dressler A., Richstone D.O., 1990, ApJ, 348, 120
%
\bibitem[Duncan \& Wheeler 1980]{duncan}
Duncan M.J. and Wheeler J.C., 1980, ApJ, 237, L27
%
\bibitem[Ferrarese et al. 1996]{ferrarese}
Ferrarese L., Ford H.C., Jaffe W., 1996, AJ, 470, 444
%
\bibitem[Ford et al. 1994]{ford94}
Ford H.C., Harms R.J., Tsvetanov Z.I., Hartig G.F., Dressel L.L., 
Kriss G.A., Bohlin R.C., Davidsen A.F.,
Margon B., Kochhar A.K., 1994, ApJ, 435, L27
%
\bibitem[Ford et al. 1996]{ford96}
Ford H.C., Tsvetanov Z.I., Hartig G.F., Kriss G.A., Harms R.J., 
Dressel L.L., 1996, ``Science with the HST -- II'',
Eds. P. Benvenuti, F. Macchetto and E. Schreier, p. 192
%
\bibitem[Harms et al. 1994]{harms}
Harms R.J., Ford H.C., Tsvetanov Z.I., Hartig G.F., Dressel L.L., 
Kriss G.A., Bohlin R.C., Davidsen A.F., Margon B., Kochhar A.K., 
1994, ApJL, 435, L35
%
\bibitem[Kormendy \& Richstone 1995]{kormendy}
Kormendy, J. and Richstone, D., 1995, ARA\&A, 33, 581
%
\bibitem[Kormendy et al. 1996]{kormendy97}
Kormendy J., et al., 1996, ApJ, 459, L57
%
\bibitem[Jaffe et al. 1993]{jaffe}
Jaffe W., Ford H.C., Ferrarese, L., van den Bosch, F.C., O'Connell R.W., 1993,
Nature, 364, 214
%
\bibitem[Jarvis \& Peletier 1991]{jarvis}
Jarvis M., Peletier R.F., 1991, A\&A, 247, 315
%
\bibitem[Macchetto 1996]{macchetto96}
Macchetto F., 1996, ``Science with the HST -- II'', Paris, Eds. P. Benvenuti,
F. Macchetto and E. Schreier, p. 59
%
\bibitem[Macchetto et al. 1997]{bigpaper}
Macchetto F.D., Marconi A., Axon D.J., Capetti A., Sparks W., Crane P.,
1997, ApJ, submitted
%
\bibitem[Nota, Jedrzejewski \& Hack, 1995]{focman}
Nota A., Jedrzejewski R., Hack W., 1995,
{\sl Faint Object Camera Instrument Handbook Version
6.0}, Space Telescope Science Institute
%
\bibitem[Rees 1997]{rees}
Rees M. J. , 1997, in ``Unsolved problems in astrophysics'',
Princeton University Press, eds. J.N.  Bachall and J.P. Ostriker, 181
%
\bibitem[Sargent 1978]{sargent}
Sargent W.L.W., Young P.J., Boksenberg A., Shortridge K., Lynds C.R.,
Hartwick F.D.A., 1978, ApJ, 221, 731
%
\bibitem[Sparks et al. 1996]{sparks}
Sparks, W.B., Biretta, J.A. and Macchetto F., 1996, ApJ, 473, 254
%
\bibitem[van der Marel 1994]{vandermar94}
van der Marel R.P., 1994, MNRAS, 270, 271
%
\bibitem[van der Marel et al. 1997a]{vandermar97}
van der Marel R.P., de Zeeuw P.T., Rix H.W., Quinlan G.D.,
1997a, Nature, 385, 610
%
\bibitem[van der Marel et al. 1997b]{vandermar97s}
van der Marel R.P., de Zeeuw P.T., Rix H.W., 1997b, ApJ submitted
%
\bibitem[Whittle et al. 1988]{whittle}
Whittle, D. M., Ward, M. J., Meurs, E. J. A., Pedlar, A., Unger, S. W. 
and  Axon, D. J., 1988, ApJ, 326, 125
%
\bibitem[Wilkins \& Axon 1992]{figaro}
Wilkins T.W. and Axon D.J., 1992, in Astronomical data analysis
software and systems I, Ast. Soc. Pac. Conf. Ser. 25, p. 427
%
\bibitem[Young et al. 1978]{young}
Young P.J., Westphal J.A., Kristian J., Wilson C.P., Landauer F.T.,
1978, ApJ, 221, 721
%
\bibitem[Young 1980]{young80}
Young P.J., 1980, ApJ, 242, 1232

\end{thebibliography}
\end{document}